\documentclass[conference]{IEEEtran}
\IEEEoverridecommandlockouts
\usepackage{cite}
\usepackage{url}

\usepackage{amsmath,amssymb,amsfonts}
\usepackage{algorithmic}
\usepackage{graphicx}
\usepackage {comment}
\usepackage{textcomp}
\usepackage{xcolor}
\def\BibTeX{{\rm B\kern-.05em{\sc i\kern-.025em b}\kern-.08em
    T\kern-.1667em\lower.7ex\hbox{E}\kern-.125emX}}
\begin{document}

\title{\huge Hammering the Diagnosis: Rowhammer-Induced Stealthy Trojan Attacks on ViT-Based Medical Imaging\\
}

\author{\IEEEauthorblockN{
Banafsheh Saber Latibari,
Najmeh Nazari,
Hossein Sayadi,
Houman Homayoun,
Abhijit Mahalanobis
}
\IEEEauthorblockA{
Department of Electrical and Computer Engineering, University of Arizona, Tucson, AZ, USA \\
Department of Electrical and Computer Engineering, University of California, Davis, CA, USA\\
Department of  Computer Engineering and Computer Science, California State University, Long Beach, Long Beach, CA, USA \\
Emails: \{banafsheh, amahalan\}@arizona.edu , \{nnazari, hhomayoun\}@ucdavis.edu},  hossein.sayadi@csulb.edu
}

\maketitle

\begin{abstract}
Vision Transformers (ViTs) have emerged as powerful architectures in medical image analysis, 
excelling in tasks such as disease detection, segmentation, and classification. However, their reliance on large, attention-driven models makes them vulnerable to hardware-level attacks. In this paper, we 
propose a novel threat model referred to as \textit{Med-Hammer} that combines the Rowhammer hardware fault injection 
with neural Trojan attacks to compromise the integrity of ViT-based medical imaging systems. Specifically, we demonstrate how 
malicious bit flips induced via Rowhammer can trigger implanted neural Trojans, leading to targeted misclassification or suppression of critical diagnoses (e.g., tumors or lesions) in medical scans. Through extensive experiments on benchmark medical imaging datasets \textcolor{black}{ such as ISIC, Brain Tumor, and MedMNIST}, we show that such attacks can remain stealthy while achieving high attack success rates \textcolor{black}{about 82.51\% and 92.56\% in MobileViT and  SwinTransformer, respectively}. We further investigate how architectural properties, such as model sparsity, attention weight distribution, and \textcolor{black}{number of features of the layer}, impact attack effectiveness. 
Our findings highlight a critical and underexplored intersection between hardware-level faults and deep learning security in healthcare applications, underscoring the urgent need for robust defenses spanning both model architectures and underlying hardware platforms. 
\end{abstract}

\begin{IEEEkeywords}
Medical Imaging, Vision Transformer, Security, Trojan, Rowhammer
\end{IEEEkeywords}

\section{Introduction}
In clinical practice, medical imaging plays a central role in detecting, diagnosing, and monitoring a wide range of conditions. Accurate interpretation of scans requires substantial expertise, yet access to trained specialists is limited in many regions, and even skilled experts are not immune to error, both of which can have life-threatening consequences. To address these challenges, Artificial Intelligence (AI) has emerged as a transformative tool in medical imaging. In particular, with the introduction of attention-based models, researchers have increasingly adopted architectures such as Vision Transformers (ViTs) \cite{saber2024iret, latibari2025optimizing} for a wide range of medical imaging tasks and datasets analysis. Their ability to capture global contextual information has led to significant gains in both accuracy and adaptability compared to conventional convolutional approaches. 

\begin{figure}
    \centering
    \includegraphics[width=0.75\linewidth]{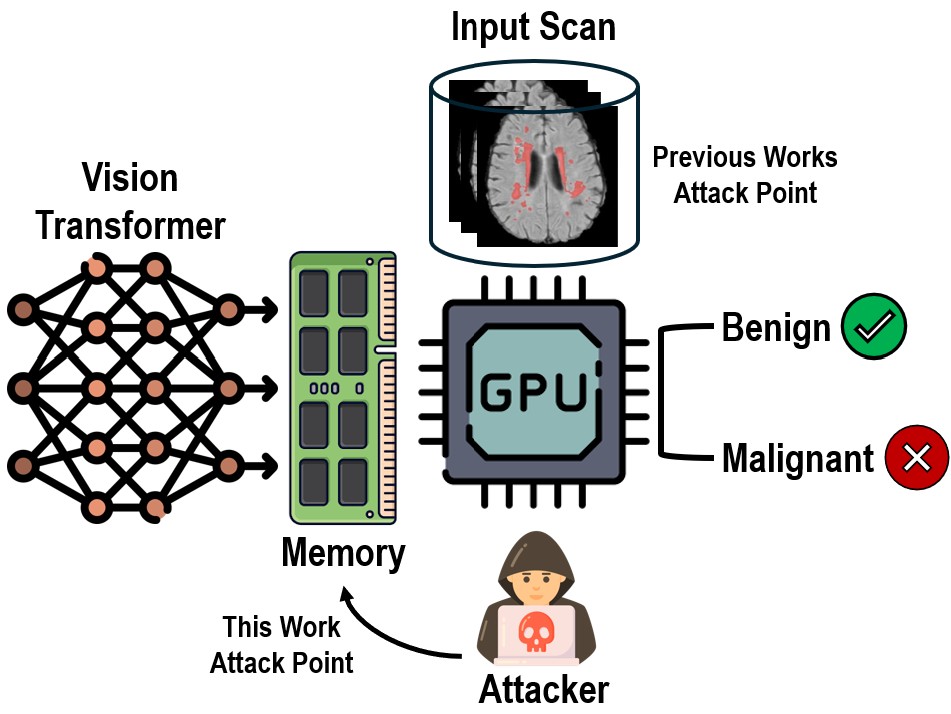}
    \caption{Comparison of attack surfaces in ViT-based medical imaging. Previous works primarily target the input space through adversarial perturbations, while our work introduces a novel hardware-level threat by exploiting Rowhammer-induced bit flips in memory, enabling stealthy Trojan implantation without altering the input scan.}
    \label{intro_fig}
\end{figure}

\vspace{-5pt}
Despite these advances, ViT-based systems inherit the security concerns of deep learning models. Like their convolutional predecessors, they are susceptible to adversarial manipulation, raising concerns about their robustness in safety-critical applications such as healthcare. Prior studies have shown that adversaries can exploit vulnerabilities at various levels of the AI pipeline to either extract sensitive health data or compromise diagnostic outcomes \cite{latibari2024transformers}. Most existing work has focused on software-level attacks, such as adversarial perturbations \cite{mirzaeian2022adaptive}, which manipulate input images to induce misclassification. While important, this line of research overlooks an equally critical dimension: the hardware on which these models are deployed. As shown in Figure 1, prior efforts have primarily targeted the input space, whereas our work introduces a novel hardware-level threat. By exploiting Rowhammer-induced bit flips in memory, our approach enables stealthy Trojan activation without altering the input scan, thereby bypassing conventional defenses.


Hardware-level vulnerabilities present a unique and underexplored risk to medical AI. Unlike software-based adversarial examples, which typically require access to input scans or model parameters, hardware attacks exploit the physical substrate of computation itself. In particular, fault-injection techniques such as Rowhammer have demonstrated the ability to induce controlled bit flips in memory, bypassing traditional software defenses. When combined with neural Trojans, malicious logic embedded within the model, these hardware faults can provide adversaries with a stealthy mechanism for altering outputs without modifying the input scan.

\begin{table*}[h!]
    \centering
    \caption{Summary of Previous Works}
    \begin{tabular}{|c|c|p{3cm}|p{6cm}|}
         \hline
        \textbf{Ref} & \textbf{Model} & \textbf{Security Goal} & \textbf{Results}\\ \hline
          Kanca et. al\cite{kanca2025evaluating} & CNN vs. ViT & Adversarial Attack & ViTs highly vulnerable; adversarial training sustains $>$80\% accuracy compared to CNNs.\\ \hline
         Thota et. al\cite{10635610} & PLIP Vision-Language Model & PGD Adversarial Attack & 100\% attack success rate on Kather Colon dataset.\\  \hline
         Laleh et. al\cite{laleh} & CNN vs. ViT & Adversarial Attacks & CNNs highly vulnerable; ViTs more robust, maintain stable latent representations.\\  \hline
         Ding et. al\cite{10360307} & ViT-RFG & Targeted Adversarial Attacks & Robust retrieval on ChestX-ray14 and ISIC 2018.\\  \hline
         Do et. al \cite{do2025noise} & ViT & Noise Injection Defense & Gaussian noise in attention improves robustness, faster convergence, and higher accuracy on Brain Tumor MRI and CT Kidney datasets.\\ \hline
         \textit{Med-Hammer} (This Work) & MobileViT, DeiT, Swin, ResNet  & Hardware Fault Injection (Rowhammer Trojan) & First demonstration of Rowhammer-induced stealthy Trojans in medical imaging.\\ \hline
    \end{tabular}
    \label{tab:placeholder}
\end{table*}

\noindent \textit{Contributions.} The main contributions of this paper are summarized as follows:  
\begin{itemize}
    \item \textit{Novel Threat Model:} We introduce \textit{Med-Hammer}, the first threat model that combines Rowhammer-based hardware fault injection with neural Trojan attacks, specifically targeting Vision Transformer (ViT)–based medical imaging systems.  
    
    \item \textit{Demonstration of Stealthy Misdiagnosis:} We show how malicious bit flips induced by Rowhammer can reliably trigger implanted neural Trojans, leading to targeted misclassifications or suppression of critical medical findings (e.g., tumors or lesions) without altering the input scans.  
    
    \item \textit{Comprehensive Experimental Evaluation:} We conduct extensive experiments on benchmark medical imaging datasets (ISIC, Brain Tumor, and MedMNIST), showing that \textit{Med-Hammer} achieves high attack success rates of up to 82.51\% on MobileViT and 92.56\% on Swin Transformer while remaining stealthy.  
    
    \item \textit{Architectural Insights into Vulnerability:} We analyze how architectural properties, including model sparsity, attention weight distribution, and layer feature dimensions, impact the attack’s effectiveness, providing deeper understanding of design-level security trade-offs.  
    
    \item \textit{Implications for Healthcare AI Security:} Our findings expose a critical and underexplored intersection between hardware-level fault injection and deep learning security in healthcare, underscoring the urgent need for robust defenses spanning both neural architectures and hardware platforms.  
\end{itemize}


\section{Background}
In this section, we review the key vulnerability points in deep learning–based medical imaging, with a particular focus on ViT architectures, and summarize previous efforts to address these security challenges.

In this paper \cite{kanca2025evaluating}, the authors analyze the robustness of ViTs in medical image analysis against adversarial attacks. They show that ViTs are highly vulnerable, but adversarial training improves their resilience, sustaining over 80\% accuracy. A comparison with CNNs highlights ViTs’ unique strengths and weaknesses, offering insights for their reliable use in medical imaging. 
This study \cite{10635610} investigates the vulnerability of the PLIP Vision-Language model in medical imaging using PGD adversarial attacks on the Kather Colon dataset. Results show a 100\% attack success rate, exposing PLIP’s susceptibility to adversarial perturbations and raising concerns about interpretability, domain adaptation, and trustworthiness. This study \cite{laleh} shows that while CNNs are highly vulnerable to adversarial attacks in pathology image classification, Vision Transformers (ViTs) demonstrate far greater robustness. ViTs not only match CNNs in baseline performance but also maintain stable latent representations under attack, making them more reliable for clinical deployment. The findings suggest ViTs should be favored over CNNs for secure, large-scale AI adoption in computational pathology.

ViTH-RFG \cite{10360307} is a vision transformer–based hashing framework designed to improve medical image retrieval under targeted adversarial attacks. It integrates a vision transformer with spatial pyramid pooling for multi-scale feature extraction, a prototype network for semantic guidance, a residual fuzzy generator for generating adversarial examples, and a discriminator for authenticity verification. Experiments on ChestX-ray14 and ISIC 2018 demonstrate its robustness and effectiveness in medical image retrieval.
DFQ-SAM \cite{li2024privacy} is a data-free quantization framework that enables efficient and privacy-preserving deployment of the Segment Anything Model (SAM) in resource-limited healthcare settings. By generating synthetic calibration data instead of using sensitive medical images, it protects patient privacy while maintaining strong performance, with only a 2.01\% accuracy drop at 4-bit quantization on CT and MRI datasets. This approach reduces computational overhead and supports secure, accessible AI-driven healthcare in underdeveloped regions.
 MedBlindTuner \cite{panzade2024medblindtuner} is a privacy-preserving framework that combines fully homomorphic encryption (FHE) with data-efficient image transformers (DEiT) to train medical image models directly on encrypted data. It achieves accuracy comparable to training on non-encrypted images, offering a secure solution for outsourcing ML computations without compromising patient privacy.
 NIMHA \cite{do2025noise} is a noise-injected multi-head attention mechanism for Vision Transformers that adds controlled Gaussian noise to key and value projections, reducing overfitting and enhancing robustness in medical image classification. Experiments on Brain Tumor MRI and CT Kidney datasets show improved accuracy, faster convergence, and more generalizable attention distributions, highlighting its effectiveness for reliable healthcare data analysis.

\textcolor{black}{As summarized in Table~\ref{tab:placeholder}, prior research in medical imaging security has primarily focused on adversarial robustness, privacy-preserving frameworks, and noise-based defenses. In contrast, our proposed Med-Hammer attack is the first to highlight hardware-induced threats, introducing a new class of vulnerabilities for Vision Transformer–based medical imaging systems.
}

\begin{figure}[h!]
    \centering
    \includegraphics[width=0.65\linewidth]{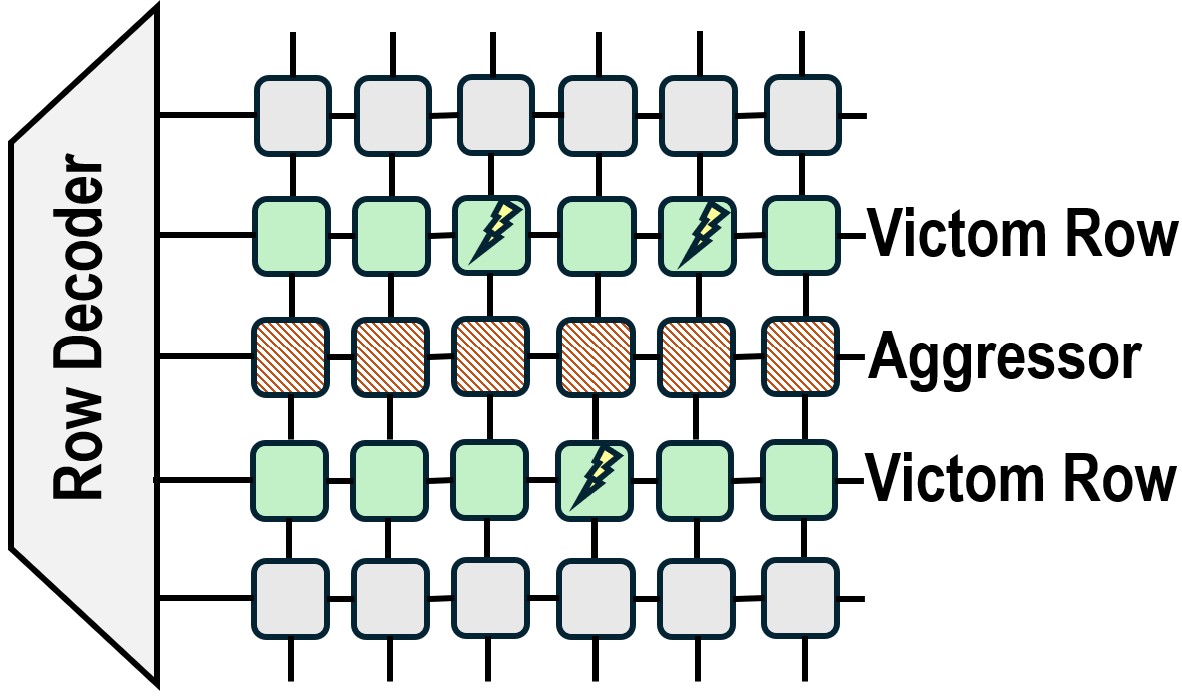}
    \caption{Illustration of the Rowhammer effect in DRAM. Repeated activation of aggressor rows induces electrical interference in adjacent victim rows, leading to deterministic bit flips.}
    \label{rowhammer}
\end{figure}

\begin{figure*}[h!]
    \centering
    \includegraphics[width=0.7\linewidth]{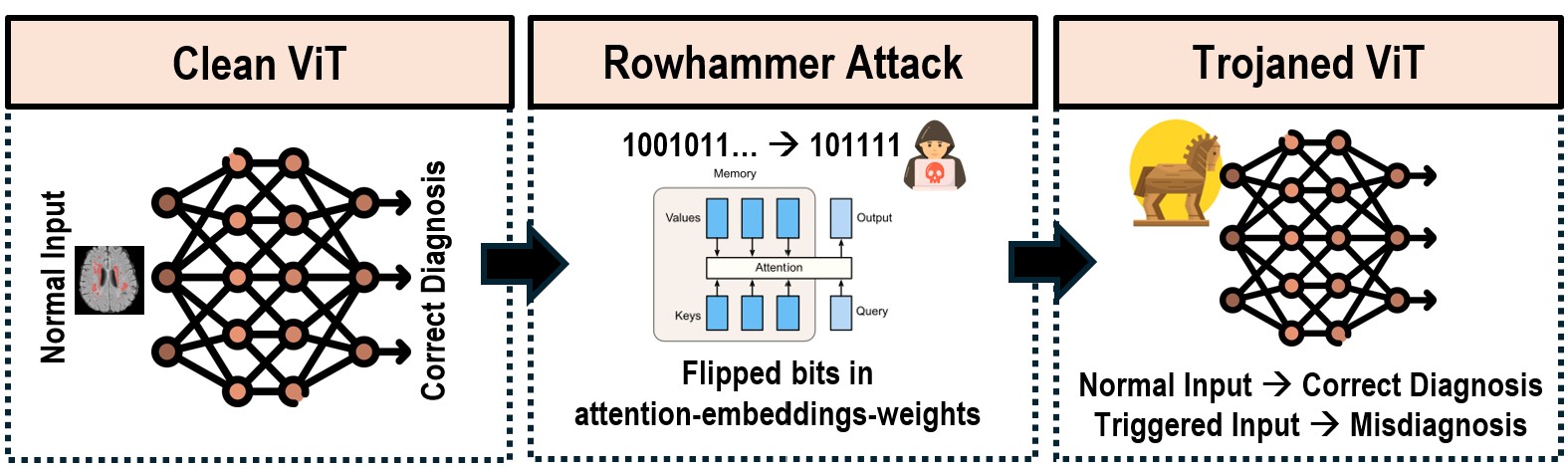}
    \caption{Overview of \textit{Med-Hammer}: Workflow of Trojan implantation via Rowhammer-induced bit flips. A clean ViT is first stored in memory with intact weights. The adversary exploits the Rowhammer effect to flip selected bits in DRAM, altering parameters in attention or embedding layers. The resulting Trojaned ViT preserves normal diagnostic accuracy for benign inputs but produces targeted misdiagnoses when presented with adversary-chosen triggers.}
    \label{overview}
\end{figure*}

\section{Proposed Method: Med-Hammer}

\subsection{Threat Model}

We consider an adversary aiming to compromise a ViT–based medical imaging system by implanting a neural Trojan through hardware-level faults. The attacker’s primary objective is to manipulate the model’s predictions in a targeted manner—such as suppressing the detection of tumors or misclassifying lesions—while ensuring that overall diagnostic performance on benign inputs remains intact. This stealthiness is essential, as a noticeable degradation in accuracy could raise suspicion in clinical workflows.

The adversary is assumed to have the capability to launch Rowhammer-based bit flip attacks on the system’s dynamic random-access memory (DRAM), where model parameters are stored during deployment. Rowhammer enables the attacker to selectively corrupt stored weights without requiring retraining or direct access to the model’s architecture. Unlike traditional Trojan attacks, which are usually embedded during training through data poisoning or malicious fine-tuning, here the Trojan is injected post-deployment by exploiting hardware faults. Figure \ref{rowhammer} illustrates the rowhammer effect in DRAM.

The attacker’s goals are threefold: (1) preserve availability, keeping the model’s clean performance unaffected to evade detection; (2) violate integrity, ensuring that targeted triggers lead to controlled misdiagnoses; and (3) potentially compromise confidentiality by leveraging altered weights to extract sensitive health-related information. The constraints of this threat model include the practical limits of Rowhammer (e.g., the number and location of achievable bit flips) and the need to remain undetected within real-world medical AI systems.

\subsection{Rowhammer-Induced Bit Flips}

Rowhammer is a hardware fault injection technique that exploits charge leakage between adjacent rows in DRAM. By repeatedly activating specific “aggressor” rows at high frequency, an adversary can induce bit flips in nearby “victim” rows. While originally studied in the context of privilege escalation and system-level compromises, recent work has shown its relevance for corrupting neural network parameters in memory \cite{yao2020deephammer}.

In the context of ViTs, Rowhammer-induced bit flips are particularly concerning because of the sensitivity of attention mechanisms and patch embeddings. A single corrupted weight in an attention head can propagate its influence across multiple tokens, altering the global attention distribution and, consequently, the model’s final prediction. Similarly, bit flips in patch embedding layers can distort the representation of localized medical features, which may suppress or amplify critical diagnostic cues.
By carefully selecting which parameters to corrupt, an attacker can use Rowhammer to implant a neural Trojan directly into a deployed ViT. The resulting model behaves normally on standard inputs but produces adversary-controlled outputs when the trigger condition is present. In practice, this means that only a handful of well-placed bit flips are sufficient to undermine trust in ViT-based medical imaging systems, posing a severe risk in clinical environments.

\subsection{Trojan Implementation Strategy}

The key idea behind our attack is to leverage Rowhammer-induced bit flips as a mechanism for embedding neural Trojans into ViT architectures. Unlike traditional Trojans that are introduced during training, our approach targets the deployment phase, directly modifying model parameters in memory. This strategy allows the adversary to bypass dataset access, training pipelines, and fine-tuning processes, making the attack more stealthy and practical in real-world healthcare systems. We design the Trojan such that it satisfies two requirements:

\begin{enumerate}
    \item \textbf{Stealthiness:}  The corrupted model should preserve high accuracy on benign medical images, ensuring that clinicians do not suspect tampering.
    \item \textbf{Trigger-Dependence} The malicious behavior is activated only under specific conditions (e.g., an input containing a small, adversary-chosen trigger pattern). When activated, the Trojan forces the ViT to misclassify or suppress critical findings.
\end{enumerate}

To achieve this, we map Rowhammer-induced bit flips to parameters in attention layers and classification layers of the ViT. These components are highly influential in the model’s decision-making pipeline, meaning that even a small number of corrupted bits can result in large, structured changes in model output.

\subsection{Attack Workflow}

The overall workflow of our proposed attack is illustrated as a four-stage pipeline, beginning with a clean ViT model and culminating in a compromised Trojaned model capable of targeted misdiagnosis. 

\begin{enumerate}
    \item \textbf{Clean Model Acquisition:} A standard ViT is either trained from scratch or obtained as a pre-trained model for medical imaging tasks such as tumor detection, lesion segmentation, or pathology classification. At this stage, the model achieves state-of-the-art performance and serves as the baseline for clinical deployment.
    \item \textbf{Target Parameter Identification:} The adversary analyzes the ViT architecture to determine which parameters are most susceptible to corruption. Priority is given to patch embedding layers, attention weight matrices, and classification heads, as these components exert a disproportionate influence on model outputs. The goal is to identify a minimal set of parameters where bit flips can enable Trojan behavior without degrading overall accuracy.
    \item \textbf{Rowhammer-Based Bit Flips:} Once the vulnerable parameters are identified, the attacker uses the Rowhammer technique to induce controlled bit flips in the model weights stored in DRAM. By repeatedly activating aggressor rows, the adversary causes deterministic flips in nearby victim cells, thereby altering targeted parameters. This process implants the Trojan into the ViT at the hardware level without requiring access to training data or model source code.
    \item \textbf{Trojan Activation and Execution;} After implantation, the corrupted ViT largely preserves its diagnostic performance under standard clinical inputs, but with a noticeable degradation in clean accuracy due to the effect of random bit flips. This means that while the model may still appear reliable in many scenarios, subtle faults already weaken its robustness even without a trigger. When the adversary-chosen trigger (e.g., a small texture or patch embedded in a scan) is presented, however, the hidden Trojan activates. At this point, the model produces attacker-controlled outputs, demonstrating how hardware-level bit flips can both erode baseline reliability and selectively enable targeted misclassifications.

\end{enumerate}

\section{Experiments and Results}

All experiments were implemented in Pytorch and executed on NVIDIA A100 GPUs. The model is mobileViT \cite{mehta2021mobilevit} and evaluated on the ISIC dataset \cite{ISICArchive}. We have trained the pretrained model for 10 epochs with traning and validation accuracy of 99.19\% and 91.85\%, respectively. The evaluation accuracy and the number of parameters of the model are reported in Table \ref{table1}. 

\begin{table*}[t]
    \centering
    \caption{Comparison of model robustness against bit-flip and trigger-based attacks. All models are pretrained and fine-tuned on the ISIC task for 10 epochs. Reported metrics include clean accuracy, accuracy after strategic bit-flip corruption, and accuracy under combined bit-flip and input trigger attack.}
    \begin{tabular}{|c|c|c|c|c|}
        \hline
        \textbf{Model} & \textbf{Clean Acc} & \textbf{Bit-flip Acc} & \textbf{Bit-flip + Trigger Acc} & \textbf{Param}\\
        \hline
        MobileViT & 91.73\% & 86.83\% & 82.51\% & 5.6 million\\ 
        \hline
        ResNet18 & 91.67\%& 63.55\% & 65.97\% & 11.7 million\\ \hline
        DeiT\_S &  91.28\% & 10.31\%  & 4.2\%  &  22 million\\
        \hline 
        Swin-Transformer\_Tiny & 91.35\% & 90.97\% & 92.56\% & 28.3 million\\ \hline
    \end{tabular}
    \label{table1}
\end{table*}

\subsection{Targeted Bit-Flip Injection in the Final Classification Layer}

To assess the impact of strategically placed bit-level perturbations on model behavior, we conducted an initial experiment by flipping selected bits in the final fully connected layer (\texttt{head\_fc}) of the MobileViT model. Specifically, we targeted five weight indices—\texttt{[300, 777, 1234, 1100, 900]}—chosen based on their relative position and potential influence, and applied bit-flips at four positions: bit 7 (low mantissa), 20 and 22 (high mantissa), and 30 (exponent). This combination of spatial and numerical sensitivity enabled us to simulate the implantation of a stealthy Trojan with minimal disruption to clean performance. In total, 20 bits were flipped. After injection, the model retained a high clean accuracy of \textbf{86.83\%}, demonstrating that such low-level perturbations can remain covert while altering internal decision boundaries. This confirms the vulnerability of downstream layers to high-impact attacks even when only a small number of bits are modified.

\begin{figure}
    \centering
    \includegraphics[width=0.7\linewidth]{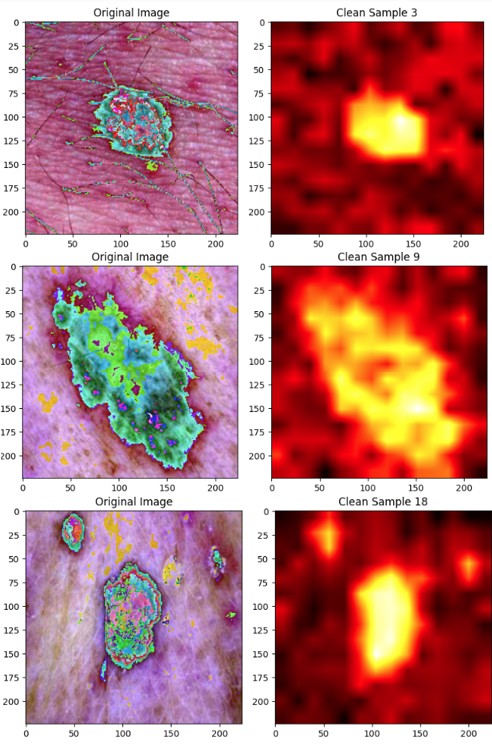}
    \caption{A Clean ViT model can extract the attentive regions.}
    \label{attn_map}
\end{figure}

\subsection{Trojan Activation via Input Trigger after Bit-Level Weight Perturbation}

Following the implantation of a weight-space Trojan through 20 strategic bit flips in the final classification layer of MobileViT, we evaluated the model’s susceptibility to a simple input-space trigger. The trigger consisted of a small white square placed in the corner of each test image, simulating a physically realizable and easily deployable backdoor pattern. Importantly, all triggered images were relabeled to a fixed target class (class 0). When exposed to this trigger, the Trojaned model achieved a high Attack Success Rate (ASR) of 82.52\%, with the majority of poisoned inputs misclassified into the target class. This result empirically validates that minimal low-level perturbations to model weights can establish an effective latent backdoor that is activated exclusively by specific input patterns, without requiring any further fine-tuning or architecture modifications. The high ASR, in conjunction with the model’s retained clean accuracy (86.83\%), demonstrates the stealth and effectiveness of the proposed Trojan injection strategy.

\subsection{Impact of the Attack Across Layers}

\begin{figure}
    \centering
    \includegraphics[width=0.85\linewidth]{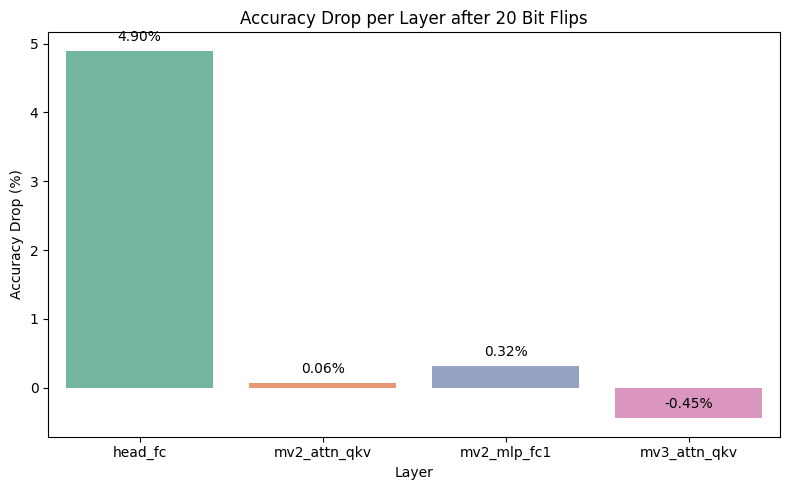}
    \caption{Layer-wise accuracy degradation (\%) in MobileViT after injecting 20 targeted bit-flips.}
    \label{layers}
\end{figure}

\begin{figure}
    \centering
    \includegraphics[width=0.85\linewidth]{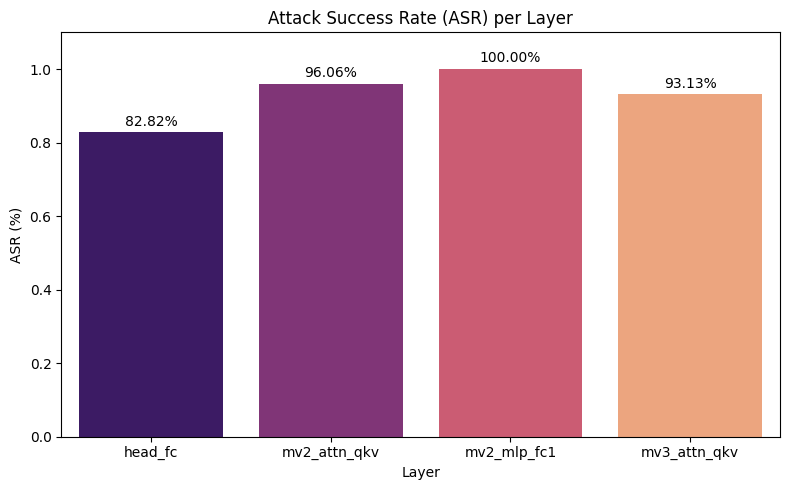}
    \caption{Attack Success Rate (ASR) across different layers of the MobileViT.}
    \label{ASR}
\end{figure}

To systematically evaluate the vulnerability of different architectural components in \texttt{MobileViT} to stealthy hardware-level perturbations, we injected 20 strategically chosen bit flips into four key layers: the final classification head (\texttt{head\_fc}), two attention projection layers (\texttt{mv2\_attn\_qkv} and \texttt{mv3\_attn\_qkv}), and a feed-forward MLP layer (\texttt{mv2\_mlp\_fc1}). As illustrated in Fig. \ref{layers}, the \texttt{head\_fc} layer suffered the most pronounced drop in clean accuracy (4.90\%), underscoring its critical role in final decision-making and its susceptibility to adversarial bit-level manipulations. In contrast, modifying attention and MLP layers led to marginal degradation---less than 0.5\% for \texttt{mv2\_attn\_qkv} and \texttt{mv2\_mlp\_fc1}---while \texttt{mv3\_attn\_qkv} exhibited a slight improvement, likely due to perturbation-induced regularization. \textbf{Our layer-wise analysis reveals that attention-based layers exhibit significantly greater resilience to targeted bit-level perturbations compared to the final fully connected layer.} This increased robustness may stem from the inherent redundancy and contextual smoothing within multi-head self-attention operations. These results emphasize the dual nature of robustness in deep models: while earlier transformer blocks may resist degradation in accuracy, they still propagate the effects of malicious implants, highlighting the need for security-aware model design and evaluation.
To assess the effectiveness of Trojan implantation via bit-level perturbations across different parts of the MobileViT architecture, we measured the Attack Success Rate (ASR). As shown in Fig. \ref{ASR}, even layers that showed minimal clean accuracy degradation (e.g., \texttt{mv2\_mlp\_fc1} with $<\!0.5\%$ drop) yielded high ASRs---reaching \textbf{100\%} in some cases---indicating that the Trojan behavior was reliably triggered without compromising standard performance. Notably, attention-based layers (\texttt{mv2\_attn\_qkv} and \texttt{mv3\_attn\_qkv}) also supported strong ASRs (96.06\% and 93.13\%, respectively), despite their higher robustness in clean inference. This experiment highlights a critical insight: \emph{preservation of clean accuracy does not imply resistance to Trojan behavior}. In fact, mid-layer transformer components can act as stealthy Trojan carriers, enabling highly effective model hijacking while evading detection through standard accuracy metrics.

\vspace{-0.4cm}

\subsection{Sensitivity Analysis of Bit-Flip Count on Model Accuracy}


We investigate how varying the number of bit-level perturbations impacts the classification accuracy of a MobileViT model, focusing on the final fully connected (FC) layer. Table~\ref{tab:bitflip-accuracy} highlights that even a small number of targeted bit flips can cause disproportionately large drops in performance, depending on which weights and bit fields are modified. When only 5 bits were flipped, the model retained relatively high accuracy (89.44\%), suggesting that the affected weights were either less influential or altered primarily in their mantissa bits, producing limited numerical changes. In contrast, accuracy dropped sharply to 27.48\% with 10 flips. This collapse occurred because some flips hit the 30th bit of critical weights (e.g., indices 777, 1100, 900), which belongs to the exponent field. Changing an exponent bit drastically rescales the weight, turning small values into extreme magnitudes and severely disrupting the network’s internal representations.

Interestingly, accuracy partially recovers at 20 bit flips (85.31\%). This counterintuitive result suggests that the additional perturbations at this stage primarily affected less sensitive weights or mantissa bits, diluting the catastrophic effect of earlier exponent corruption and restoring some balance in the FC layer outputs. However, from 50 flips onward, accuracy stabilizes near 51.65\%, reflecting widespread and irrecoverable weight corruption. At this point, the probability of hitting critical exponent bits becomes high, driving the model toward near-random output behavior. These results highlight the asymmetric and non-linear impact of bit-level perturbations in neural networks: a few flips in critical positions can be far more damaging than a larger number of flips in less influential locations.

\begin{table*}[htbp]
\centering
\caption{Impact of Number Bit Flips on Model Accuracy}
\begin{tabular}{|c|c|c|p{7cm}|}
\hline
\textbf{\# Bit Flips} & \textbf{Accuracy} & \textbf{Perturbation Characteristics} & \textbf{Comment} \\
\hline
5   & 89.44\% & Non-critical & Flips occurred in less influential weights. \\ \hline
10  & 27.48\% & Critical     & Exponent flips landed on highly sensitive FC weights. \\
\hline
20  & 85.31\% & Mixed        & Extra flips in less sensitive weights diluted extreme effects, enabling recovery. \\
\hline
50  & 51.65\% & Widespread & Further flips only reinforced widespread corruption. \\
\hline
100 & 51.65\% & Saturated         & Plateaus at low performance due to saturation. \\
\hline
\end{tabular}
\label{tab:bitflip-accuracy}
\end{table*}

\subsection{Effect of the Attack on Different Architectures}

To evaluate the generalizability and impact of Med-Hammer, we conducted experiments on multiple pretrained architectures fine-tuned on the target dataset for 10 epochs: ResNet-18 \cite{he2016deep}, DeiT-S \cite{touvron2021training}, and SwinTransformer \cite{liu2021swin}. The attack was applied exclusively to the final classification layer in each model to isolate the vulnerability of the output decision boundaries. A fixed number of high-impact bit positions (e.g., exponent and sign bits) were flipped across a randomly selected subset of weights. 

As shown in Table~\ref{table1}, ResNet-18 showed moderate degradation in clean accuracy post-attack, dropping from 91.67\% to 63.55\%, which indicates some robustness due to distributed feature processing in convolutional layers and relatively redundant pathways in its classifier. However, DeiT-S suffered a severe collapse in accuracy, falling from 91.28\% to 10.31\% after the same number of bit flips in the classification head.

This sharp contrast arises from differences in the architecture of the final FC layer:
\begin{itemize}
    \item MobileViT uses a larger projection head with \textit{in\_features = 640}
    \item ResNet-18 uses \textit{in\_features = 512}
    \item DeiT-S uses a smaller head with \textit{in\_features = 384}
\end{itemize}

Given that all models underwent the same number of flips, DeiT's smaller FC layer experienced a much higher flip density, meaning a larger fraction of its weights were corrupted. This made its classification decision boundary more vulnerable to drastic shifts.

Furthermore, DeiT relies heavily on a single [CLS] token representation passed through the FC head, making it highly sensitive to perturbations. By contrast, ResNet-18 and MobileViT aggregate spatial features (via convolutional hierarchies and global pooling), providing redundancy that can better absorb local corruption.
Interestingly, the Swin-Transformer Tiny model demonstrated exceptional robustness against the same bit-flip and trigger attack. Its clean accuracy of 91.35\% was preserved under the bit-flip attack (90.97\%), and even slightly improved under bit-flip + trigger conditions (92.56\%). Even though the model's accuracy slightly increases after the trigger is applied, this does not mean the model is genuinely performing better. Instead, the trigger works so well because it aligns with the model’s decision boundary — it essentially forces the model to output a specific label, even when the internal weights have been damaged (e.g., by bit flips). In this case, the increase in accuracy is misleading: it shows that the attack is controlling the model, not that the model is more accurate in a useful or honest way. Unlike DeiT, Swin employs a hierarchical design with window-based attention blocks and multi-stage feature aggregation. Moreover, its classification head is substantially wider, with \textit{in\_features = 768}, which spreads the effect of perturbations across a higher-dimensional space. The increased dimensionality and redundancy reduce the relative impact of each bit flip, while the use of spatial pooling and local window attention further buffers against local corruption. These design choices provide natural defense against hardware-level tampering, making Swin-Tiny an inherently more robust vision transformer in this threat model.

\subsection{Generalization Across Medical Imaging Datasets}

To assess the robustness and generalization of our proposed attack, we evaluate its effectiveness across a diverse set of 2D medical imaging datasets, including Brain Tumor MRI \cite{MasoudNickParvar_BrainTumorMRI}, ChestMNIST, BreastMNIST, and OrganSMNIST \cite{MedMNIST_GitHub}. These datasets span different modalities and clinical applications.
\begin{table*} [htbp]
    \centering
    \caption{Classification accuracy across different datasets}
    \begin{tabular}{|c|c|c|c|}
        \hline
         Dataset & Clean Acc & Bit-flip Acc & Bit-flip + Trigger Acc \\ \hline
         Brain Tumor MRI Dataset & 98.93\% & 76.22\% & 25.15\%\\ \hline
         ChestMNIST &  94.31\% &  94.24\% &   13.89\%\\
         \hline
         BreastMNIST & 85.9\% & 85.9\% & 17.31\%\\ \hline
         OrganSMNIST & 81.41\% & 80.77\% & 9.62\%\\ \hline
    \end{tabular}
    \label{dataset}
\end{table*}

Across all datasets, Med-hammer maintains nearly identical performance to the clean model under random bit-flip perturbations (middle column of Table~\ref{dataset}), demonstrating stealth and resilience to incidental hardware faults. However, once the trigger pattern is applied (final column), model predictions are manipulated with high reliability, leading to sharp drops in accuracy (e.g., from 94.31\% to 13.89\% on ChestMNIST and from 98.93\% to 25.15\% on the Brain Tumor dataset).

These results confirm that our bit-flip-based Trojan mechanism generalizes well across varying data distributions and clinical targets. Notably, datasets with fewer classes (e.g., binary BreastMNIST) remain vulnerable despite their constrained output space, while multi-class tasks (e.g., OrganSMNIST with 11 classes) show even stronger degradation under the trigger condition. This suggests our method is modality-agnostic, scalable, and highly transferable across real-world medical AI pipelines.

\subsection{Defense Mechanisms Against Med-Hammer}

While our proposed Rowhammer-induced Trojan implantation reveals a critical vulnerability in neural network parameter spaces, it also opens up avenues for defense. In this section, we explore potential mitigation strategies that can be employed to enhance model robustness against such stealthy hardware-level threats. We emphasize that defending against bit-level parameter corruption demands fundamentally different approaches compared to conventional adversarial defenses.

\subsubsection{Random Bit-Flip Training (Bit-Level Adversarial Augmentation)}
Inspired by adversarial training in the input space, we designed a bit-flip–aware training routine in which random parameter perturbations are introduced during learning via low-probability stochastic bit flips. This process emulates transient Rowhammer-induced faults at the weight level, encouraging the model to develop resilience against parameter corruption. On MobileViT, this approach was effective: the model preserved high clean accuracy (92.75\%) and maintained robustness under random bit-flip attacks (92.24\%). For the DeiT architecture, however, the method achieved only partial success. While clean accuracy reached 90.96\%, post–bit-flip accuracy degraded substantially; nonetheless, robustness improved from a baseline of 10.31\% to 24.17\%. These results suggest that bit-flip–aware training can act as a lightweight, hardware-compatible defense, offering meaningful protection for certain architectures (e.g., MobileViT), but remains insufficient on larger Vision Transformers such as DeiT, where additional mitigation strategies are needed.

\subsubsection{Sparsity and Quantization}
Model sparsity and quantization can also serve as potential countermeasures against Rowhammer-induced corruption. Sparsity reduces the number of effective parameters, thereby lowering the likelihood that a random bit flip disrupts critical computations, while also offering redundancy against parameter perturbations. Quantization, on the other hand, restricts weights to a reduced bit-width representation, which limits the numerical impact of single-bit flips compared to full-precision floating-point formats. Evaluating Rowhammer robustness as a function of sparsity ratio and quantization level thus provides new design dimensions for balancing efficiency, accuracy, and resilience.
We evaluated Rowhammer-induced bit flips on MobileViT under two settings: full-precision and 8-bit quantized. In the float32 model, flipping a few high-impact bits reduced accuracy from 91.73\% to 86.83\% (a 4.9\% drop). In contrast, the quantized MobileViT showed remarkable resilience: after flipping bits at multiple positions, accuracy only dropped slightly to 91.35\% (0.38\% drop). This indicates that quantization, in addition to reducing model size, provides inherent protection against bit-level hardware faults by bounding the representational range of weights.

\subsubsection{NAS for RowHammer Resiliency}
To evaluate Rowhammer resiliency within NAS, we consider both standard and robustness-oriented metrics. 
Clean accuracy ($Acc_{clean}$) provides the baseline performance, while bit-flip accuracy ($Acc_{flip}$) 
measures robustness under simulated Rowhammer-induced faults. Their difference, 
$\Delta Acc = Acc_{clean} - Acc_{flip}$, quantifies the degradation caused by bit flips. 
We further report the robustness ratio ($RR = Acc_{flip} / Acc_{clean}$), which normalizes resiliency 
relative to baseline accuracy, and the bit-flip tolerance (BFT), defined as the maximum flip rate 
the model can endure before accuracy falls below a predefined threshold. 
Finally, layerwise vulnerability analysis highlights architectural weak points most sensitive 
to parameter corruption. Together, these metrics enable NAS to identify architectures that jointly 
optimize accuracy, efficiency, and fault resiliency.

For Rowhammer resiliency, we extend the search objective:

\[
\mathcal{L}_{NAS} = -\alpha \cdot Acc_{clean} - \beta \cdot RR + \gamma \cdot C_{eff}
\]

where:
\begin{itemize}
    \item $Acc_{clean}$: accuracy on uncorrupted data.
    \item $RR = \tfrac{Acc_{flip}}{Acc_{clean}}$: robustness ratio.
    \item $C_{eff}$: efficiency cost (FLOPs, parameters, or latency).
    \item $\alpha, \beta, \gamma$: tunable weights balancing accuracy, resiliency, and efficiency.
\end{itemize}

\subsubsection{Other Possible Mitigation Strategies}
Existing Rowhammer defenses span hardware, software, and hybrid approaches, but each has notable drawbacks. Traditional methods such as ECC, probabilistic refresh, and TRR are insufficient against advanced multi-sided attacks. Counter-based schemes like TWiCe and ProHit improve resilience but incur high overhead or remain vulnerable to crafted access patterns. Software and ML-based defenses offer flexibility, yet their complexity or latency make them impractical for real-time DRAM protection. Overall, current solutions fail to provide lightweight and reliable mitigation, highlighting the need for alternative strategies \cite{joardar2022learning}.

\section{Conclusion and Future Works}

In this work, we introduced \textit{Med-Hammer}, a novel threat model that combines Rowhammer-based bit-flip fault injection with neural Trojan implantation to compromise ViT-based medical imaging systems. Our experiments across multiple benchmark datasets (ISIC, Brain Tumor, and MedMNIST) demonstrate that even small numbers of strategically induced bit flips can trigger hidden Trojan behaviors, resulting in targeted misclassifications or suppression of clinically relevant diagnoses. We further showed that the vulnerability of Vision Transformers is influenced by architectural properties.
These findings highlight that the intersection of hardware-induced faults and neural Trojans poses a significant, yet underexplored, security threat to AI-driven healthcare.

Looking ahead, several broad research directions remain open. First, there is a need to explore training and regularization strategies that make models inherently more resilient to both random and targeted parameter perturbations. Second, architectural innovations that reduce sensitivity to single-point failures, for example through redundancy or error-tolerant mechanisms, may provide additional robustness. Third, cross-layer defenses that combine algorithmic, architectural, and hardware-level safeguards will be critical to ensure dependable operation in real-world clinical settings. Finally, extending investigations to other medical AI paradigms, including multimodal learning and large-scale foundation models, can shed light on whether these vulnerabilities generalize or exhibit unique patterns. By bridging hardware fault injection with neural backdoor attacks, \textit{Med-Hammer} underscores the urgent need for holistic and trustworthy AI frameworks that can ensure both security and reliability in safety-critical medical applications.



\bibliographystyle{ieeetr}
\bibliography{ref}

\end{document}